# Comparison of Error Estimations by DERs in One-Port S and SLO Calibrated VNA Measurements and Application

*Nikolitsa YANNOPOULOU[1], Petros ZIMOURTOPOULOS[2]*

[1] Antennas Research Group, Palaia Morsini, Xanthi, Thrace, Hellas, EU
[2] Antennas Research Group, Dept. of Electrical Engineering and Computer Engineering, Democritus University of Thrace,
V. Sofias 12, 671 00 Xanthi, Thrace, Hellas, EU

yin@antennas.gr,  pez@antennas.gr, www.antennas.gr

**Abstract.** *In order to demonstrate the usefulness of the only one existing method for systematic error estimations in VNA (Vector Network Analyzer) measurements by using complex DERs (Differential Error Regions), we compare one-port VNA measurements after the two well-known calibration techniques: the quick reflection response, that uses only a single S (Short circuit) standard, and the time-consuming full one-port, that uses a triple of SLO standards (Short circuit, matching Load, Open circuit). For both calibration techniques, the comparison concerns: (a) a 3D geometric representation of the difference between VNA readings and measurements, and (b) a number of presentation figures for the DERs and their polar DEIs (Differential Error Intervals) of the reflection coefficient, as well as, the DERs and their rectangular DEIs of the corresponding input impedance. In this paper, we present the application of this method to an AUT (Antenna Under Test) selected to highlight the existence of practical cases in which the time consuming calibration technique results a systematic error estimation stripe including almost all of that of quick calibration.*

## 1. Introduction

The systematic error in a full one-port calibrated VNA measurement $\rho$ of a given one-port DUT (Device Under Test) is already estimated by its DER [1]:

$$\rho = (m - D)/[M(m - D) + R], \qquad (1)$$

$$d\rho = [-RdD - (m - D)^2 dM - (m - D)dR + Rdm] / [M(m - D) + R]^2 \qquad (2)$$

where *m* is the VNA complex reading and *D*, *M* and *R* are the complex system errors of Fig. 1.

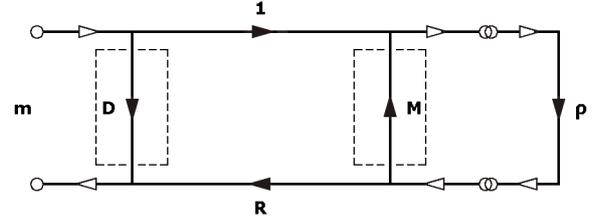

**Fig. 1.** Full one-port error model.

The relations holding between this complex reflection coefficient $\rho$ and its respective impedance *Z*, as well as, between their DERs are [1]:

$$Z = Z_0(1 + \rho)/(1 - \rho), \qquad (3)$$

$$dZ = 2Z_0 d\rho/(1 - \rho)^2 \qquad (4)$$

In this paper, we express the DERs for systematic error estimation in VNA measurements calibrated by the much simpler and quicker reflection response technique, in order to be in place to make some practical decisions from the different calibration techniques comparison.

## 2. Response Calibration

The reflection response calibration technique can be accomplished with the measurement of only one standard load, instead of three in full one-port, usually of a S short circuit [2]. This means that the flow graph of Fig. 1 is simplified a lot, since the two surrounded by dashed boxes system error branches of directivity *D* and source match *M* do not exist, equivalently $D = 0$ and $M = 0$, and (1) results to:

$$R = m/\rho_s = s/S \qquad (5)$$

where *s* is the VNA complex reading of the *S* short circuit standard with a nominal value of $S = -1$, *m* is the complex reading of a given DUT and $\rho_s$ is its complex reflection coefficient as it is measured after this response calibration:

$$\rho_s = (m/s)S \qquad (6)$$

which, from (2), has the differential error:

$$d\rho_s = (S/s)dm - (Sm/s^2)ds + (m/s)dS. \quad (7)$$

The corresponding total DER is then the sum of $L = 3$ parallelograms. Therefore, this DER contour is a polygonal line with $4L = 12$ line segments and vertices at most, in contrast with the DER of the measurement after a SLO full one-port calibration, which is a piecewise curve composed of $4(L - 1) = 24$ line segments, $4(L - 1) = 24$ circular arcs and $8(L - 1) = 48$ vertices, at most [1].

## 3. Application Results

By following the error estimation process, we already detailed in [1], we take as $dS$ the considered manufacturers' standard $S$ uncertainty data:

$$-0.01 \le d|S| \le 0, \quad -2° \le d\angle S \le +2°$$

and as $dm$ and $ds$ the VNA inaccuracy of ±1 digit in LSD of their corresponding readings, for either the amplitude in decibels or the phase in degrees. Moreover, the one-port DUT that was considered is the same typical UHF ground-plane antenna (that is: AUT) mentioned in [1].

The difference between the 3 nominal values (−1, 0, 1) of the 3 full one-port calibration standards ($S$, $L$, $O$), respectively, and their 3 corresponding VNA readings ($s$, $l$, $o$), can be estimated by the extent of the surfaces shown in the triptych of Fig. 2, where the vertical axis segment represents the range of the distinct stepped frequencies. Each surface is formed by parallel to horizontal plane lines. Each such line expresses the complex difference between the standard nominal value and its corresponding VNA reading, in each stepped frequency.

In the triptych of Fig. 3, and from left to right we have the difference surfaces made by distance lines between: (a) the measured reflection coefficient $\rho$ after a full SLO one-port calibration (black solid points) and the corresponding VNA readings $m$ for the AUT measurement (colored magenta points), (b) the measured reflection coefficient $\rho_s$ after $S$ response calibration (black ring points) and the corresponding VNA readings $m$ for the AUT measurement (colored magenta points), and (c) the two measurements ($\rho$, $\rho_s$).

All the involved, previously shown, quantities are projected on the horizontal complex plane of Fig. 4. The magenta colored spiral represents $m$, while, the black curves the reflection coefficient: solid points, for $\rho$, and ring points for $\rho_s$. All of $l$ VNA readings are close enough to complex origin (colored green points). It is rather difficult to distinguish the two curves for $s$ and $o$ VNA readings, which are close enough to the unit circle circumference (colored red solid points and colored blue ring point, respectively).

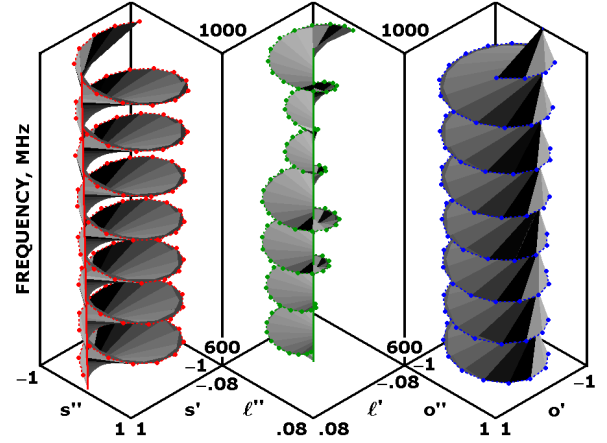

**Fig. 2.** Difference between $s$ and $S$, $l$ and $L$, $o$ and $O$.

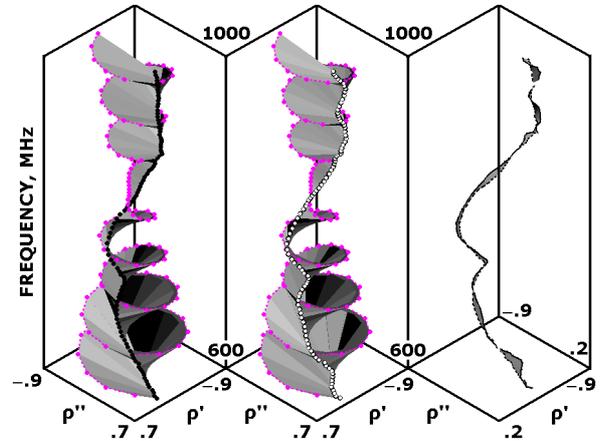

**Fig. 3.** Difference between $m$ and $\rho$, $m$ and $\rho_s$, $\rho$ and $\rho_s$.

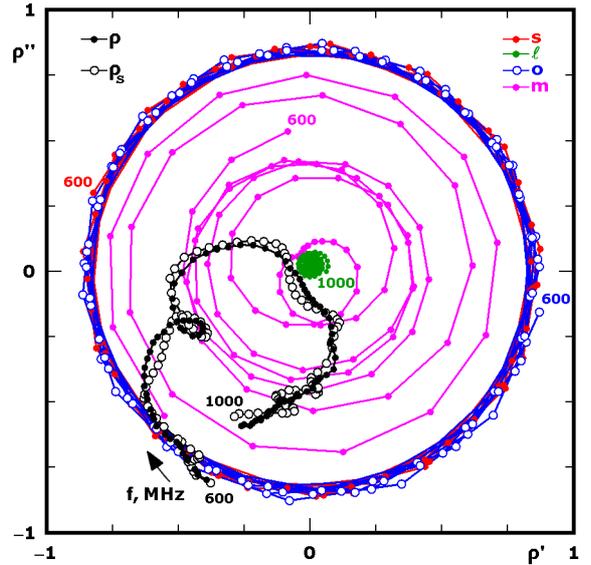

**Fig. 4.** VNA $s$, $l$, $o$, $m$ readings and $\rho$, $\rho_s$ measurements.

The $\rho$–DERs and $\rho_s$–DERs, for all 4 MHz stepped frequencies covering the range of [600, 1000] MHz, are overlapped on the complex plane of Fig. 5, forming a light and a dark gray stripes, respectively. From each stripe we selected 11 DERs out of 101, drawn with dark gray and white colors respectively, to illustrate their outline dependence on frequency.

The comparison between the AUT measurements based on these two calibration techniques is extended to the comparison against the frequency: (a) of the computed polar DEIs of the reflection coefficient magnitude and argument stripes in Fig. 7, (b) of the rectangular DEIs for the corresponding R input resistance and X input reactance stripes, in Fig. 8 and (c) of the $Z$–DERs, and $Z_s$–DERs stripes in Fig. 9.

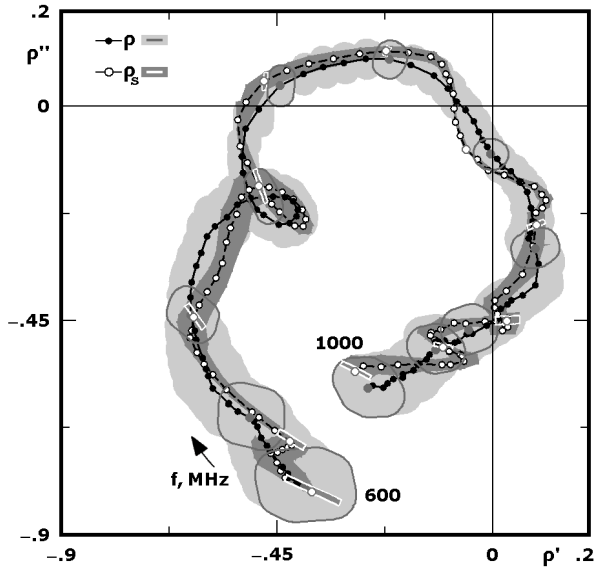

**Fig. 5.** Complex $\rho$–DERs and $\rho_s$–DERs in [600, 1000] MHz.

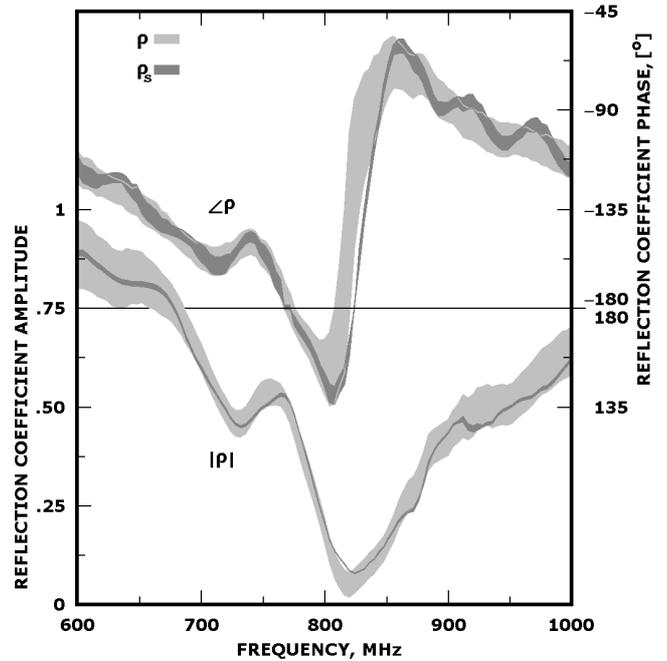

**Fig. 7.** Polar DEIs of reflection coefficient.

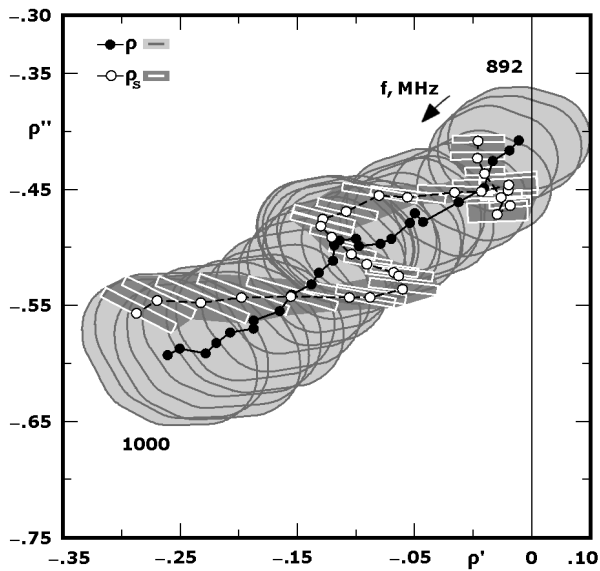

**Fig. 6.** Complex $\rho$–DERs and $\rho_s$–DERs in [892, 1000] MHz.

Moreover, we selected to magnify a part of this figure in the sub-range of [892, 1000] MHz, to further illustrate the DER outlines and their overlapping in Fig. 6, where the clearly shown ripple of the simple response calibration stripe over the relatively smooth full one-port calibration stripe reveals the superiority of the latter in the production of more accurate measurements.

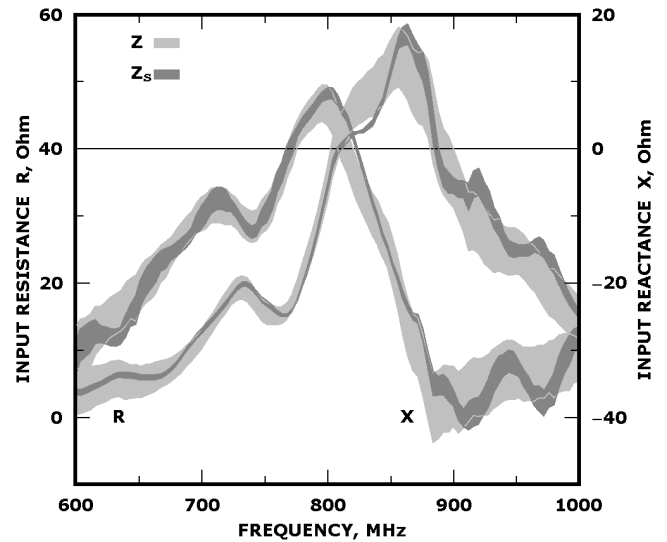

**Fig. 8.** Rectangular DEIs of input impedance.

From all that, it must be clear now that in this intentionally selected for presentation particular AUT case there was no advantage at all in selection of full one-port calibration over the reflection response one due to their remarkable in all aspects coincidence. Of course this is just another one conclusion a-posteriori.

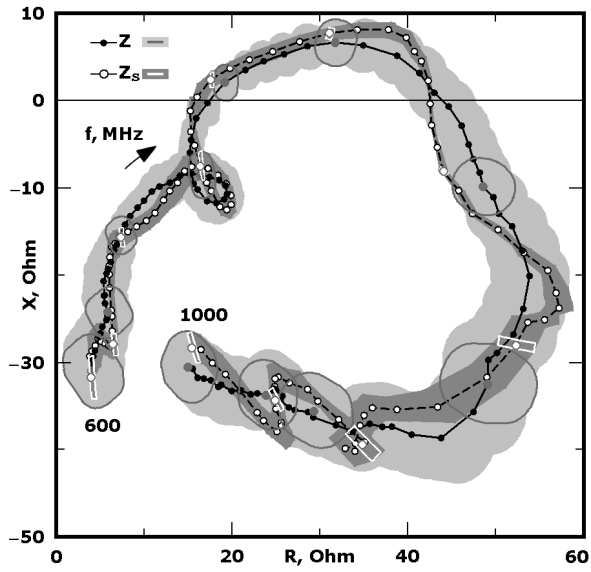

**Fig. 9.** Complex $Z$–DERs and $Z_s$–DERs in [600, 1000] MHz.